\documentclass[a4paper, 11pt]{article}
\pdfoutput=1

\usepackage{jinstpub} 

\usepackage{hyperref}
\usepackage{cleveref}
\usepackage{graphicx}
\usepackage{amssymb}
\usepackage{amsmath}

\title{Impact of a cryogenic baffle system on the suppression of radon-induced background in the KATRIN Pre-Spectrometer}

\author[a]{S.~G\"orhardt}
\author[a,d,1]{J.~Bonn,%
                \note{deceased}}
\author[a]{L.~Bornschein}
\author[a]{G.~Drexlin}
\author[a]{F.M.~Fr\"ankle}
\author[a]{R.~Gumbsheimer}
\author[a,e]{S.~Mertens}
\author[b]{F.R.~M\"uller}
\author[a]{T.~Th\"ummler}
\author[a]{N.~Wandkowsky}
\author[c]{C.~Weinheimer}
\author[b,2]{J.~Wolf,%
        \note{Corresponding author}}
\emailAdd{joachim.wolf@kit.edu}

\affiliation[a]{IKP, Karlsruhe Institute of Technology (KIT), Hermann-von-Helmholtz-Platz 1, 76344 Eggenstein-Leopolds\-hafen, Germany}
\affiliation[b]{ETP, Karlsruhe Institute of Technology (KIT), Wolfgang-Gaede-Str. 1, 76131 Karlsruhe, Germany}
\affiliation[c]{Westf\"alische Wilhelms-Universit\"at, Schlossplatz 2, 48149 M\"unster, Germany}
\affiliation[d]{Johannes Gutenberg-Universit\"at Mainz, Saarstra\ss e 21, 55122 Mainz, Germany}
\affiliation[e]{Max-Planck-Institut f\"{u}r Physik, F\"{o}hringer Ring 6, 80805 M\"{u}nchen, Germany}

\date{\today}

\abstract{
The KATRIN experiment will determine the effective electron anti-neutrino mass with a sensitivity of 
$200\,\mathrm{meV/c}^2$ at 90\% CL. The energy analysis of tritium $\beta$-decay electrons will be performed by 
a tandem setup of electrostatic retarding spectrometers which have to be operated at very low background levels of 
$<10^{-2}\,$counts per second. This benchmark rate can be exceeded by background processes resulting from the 
emanation of single ${}^{219,220}$Rn atoms from the inner spectrometer surface and an array of non-evaporable 
getter strips used as main vacuum pump. Here we report on a the impact of a cryogenic technique to reduce this radon-induced 
background in electrostatic spectrometers. It is based on installing a liquid nitrogen cooled copper baffle in the 
spectrometer pump port to block the direct line of sight between the getter pump, which is the main source of 
${}^{219}$Rn, and the sensitive flux tube volume. This cold surface traps a large fraction of emanated radon atoms 
in a region outside of the active flux tube, preventing background there. We outline important baffle design criteria to 
maximize the efficiency for the adsorption of radon atoms, describe the baffle implemented at the KATIRN Pre-Spectrometer 
test set-up, and report on its initial performance in suppressing radon-induced background.
}


\begin{document}

\maketitle
\flushbottom

\section{Introduction}
\label{sec:instruction}
 
The KATRIN (\textbf{KA}rlsruhe \textbf{TRI}tium \textbf{N}eutrino) experiment \cite{KATRINDesignReport} is 
currently being commissioned at the Karlsruhe Institute of Technology (KIT) in Germany. The experiment will 
investigate the kinematics of the tritium $\beta$-decay close to the kinematic endpoint of $E_{0} \approx 18.6\,$keV 
in a model-independent way, relying on energy and momentum conservation only. It will determine the effective 
electron anti-neutrino mass $m_{\nu}$ with a sensitivity of $200\,\mathrm{meV/c}^2$ at 90\% confidence level after 
three live years of measurements (equivalent to five calender years) \cite{drexlin2013current}. The imprint of a 
non-zero neutrino mass is a distortion of the shape of the electron energy spectrum a few eV below the 
endpoint. Due to the rather low signal countrate of a few times $10^{-2}\,$counts per second (cps) in this narrow 
region-of-interest, a low background rate of $\leq10^{-2}\,$cps has to be achieved to obtain a good signal to 
background ratio.
 
\begin{figure}[t]
  \centering
  \includegraphics[width=\textwidth]{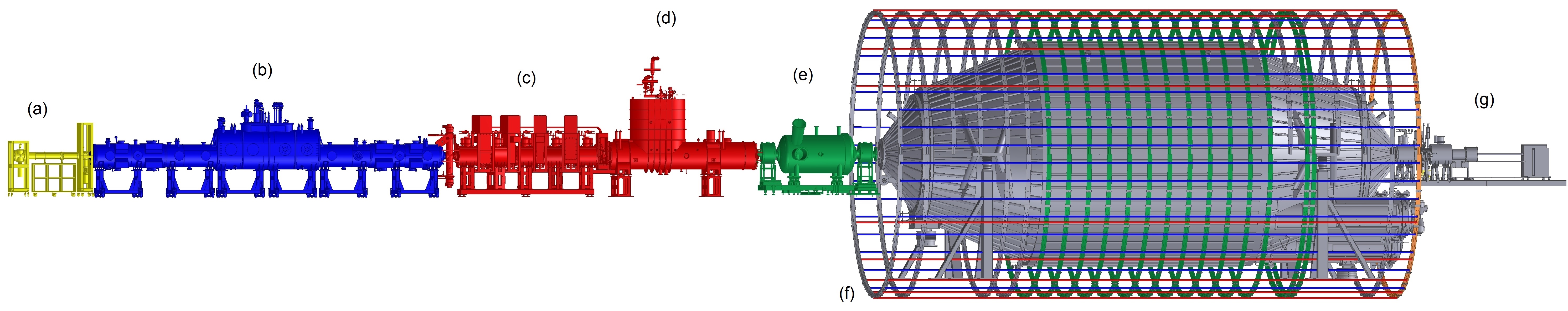}
  \caption[KATRIN setup]{Schematic overview of the $70\,\mathrm{m}$ long KATRIN setup: \textbf{a} rear section, 
\textbf{b} windowless gaseous molecular tritium source (WGTS), \textbf{c} differential pumping section (DPS), 
\textbf{d} cryogenic pumping section (CPS), \textbf{e} Pre-Spectrometer, \textbf{f} Main Spectrometer, 
\textbf{g} detector system.}
  \label{fig:beamline}
\end{figure}

A schematic overview of the $70\,\mathrm{m}$ long KATRIN setup is given in Fig.~\ref{fig:beamline}. A windowless 
gaseous tritium source (WGTS) \cite{WGTS} will provide an unprecedented activity of $\sim 10^{11}$ $\beta$-decays 
per second. Only electrons emitted in the WGTS into the forward cone are guided adiabatically by strong magnetic 
fields ($\sim 5\,$T) through the transport section to the spectrometers. The electrons are constrainted to their magnetic flux tube throughout the experiment. The two transport elements, the differential 
pumping section (DPS) \cite{Kosmider:2012:PHD} and the cryogenic pumping section (CPS) \cite{CPS}, will remove 
residual tritium gas by turbomolecular pumps (DPS) and cryotraps (CPS). The spectrometer section, consisting of a 
Pre- and a larger Main Spectrometer, analyzes the electron energy. The Pre-Spectrometer offers the option to 
act as a filter, where only electrons with energies close to $E_0$ would be transmitted to the Main Spectrometer 
for precision energy filtering, thereby limiting the incoming flux of electrons. Both spectrometers are operated as 
MAC-E filters (\cite{Bea80}, \cite{Lobashev1985}, and \cite{picard92}).

\begin{figure} [t]
  \centering
  \includegraphics[width=\textwidth]{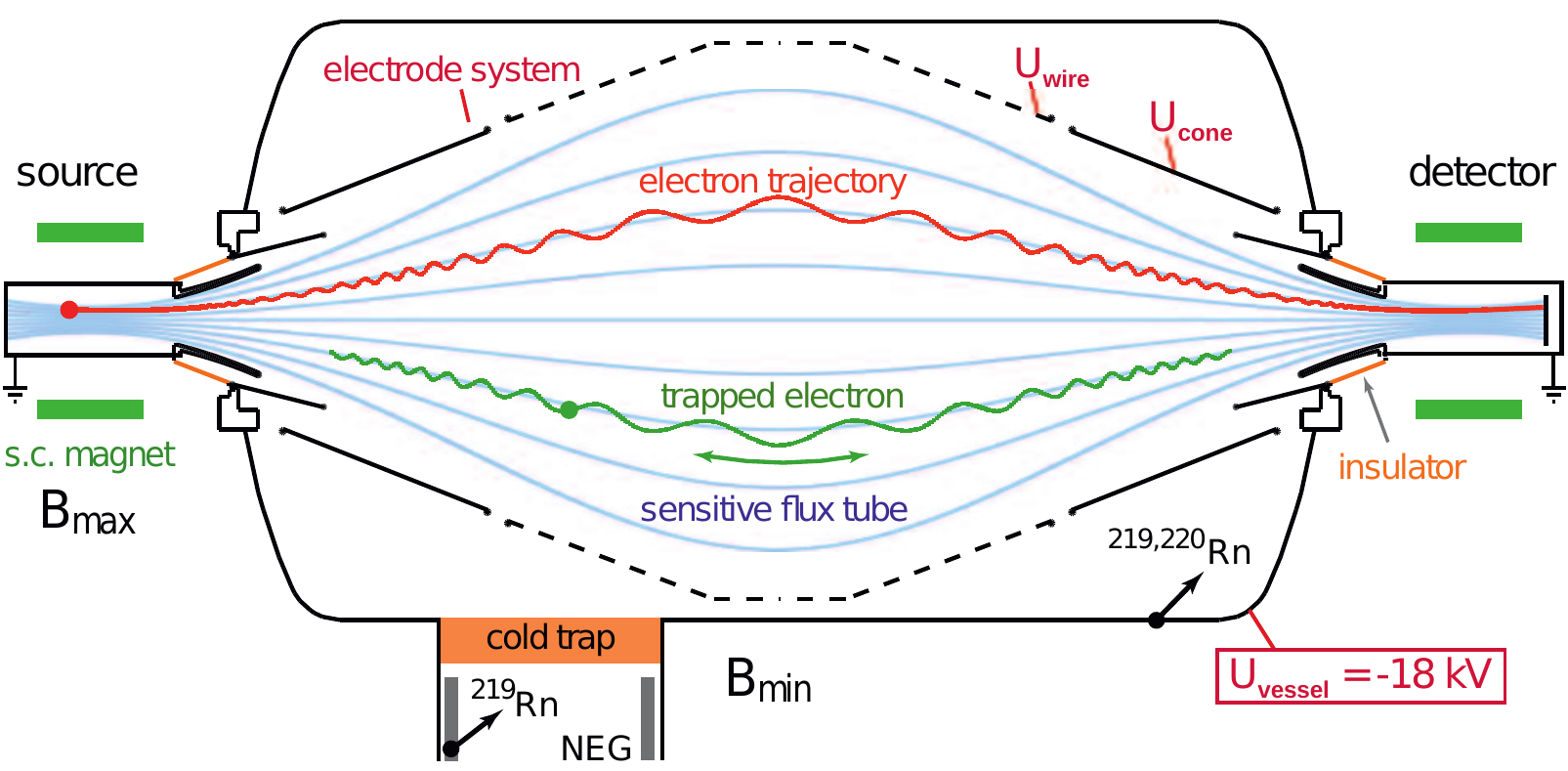}
  \caption[the pre-spectrometer MAC-E filter]{
Electromagnetic layout of the Pre-Spectrometer with two superconducting magnets providing the characteristic flux tube of the magnetic field (blue lines) that guides electrons (red line) from the source to the detector through the electro-static field of the MAC-E filter. This configuration is also responsible for the inherent magnetic bottle characteristics of a MAC-E filter trapping charged particles (green line) which are produced in the low-field region. A concentric inner electrode system of thin stainless steel cones (black line) and wires (broken lines) allows to fine-tune the retarding potential.}
\label{fig:setup}
\end{figure}

In this filter technology, the magnetic field strength in the spectrometer drops by several orders of magnitude between 
either end and the center of the vessel (see Fig.~\ref{fig:setup}). In the Main Spectrometer, this field 
ratio amounts to $1/20000$. Due to the conservation of the magnetic moment of the electrons and the adiabatic field 
layout, the magnetic gradient force transforms almost the entire cyclotron energy of the isotropically emitted electrons 
into longitudinal energy, which can then be analyzed by the electrostatic retarding potential. 
This \textbf{M}agnetic \textbf{A}diabatic \textbf{C}ollimation with an \textbf{E}lectrostatic filter, the 
MAC-E principle, allows the spectrometers to act as a high-pass filter for electrons. Accordingly, only those 
electrons that are able to overcome the potential barrier in the spectrometer section are counted by a monolithic 
segmented Si-PIN diode array \cite{Amsbaugh2015} at the downstream end of the Main Spectrometer. 
By 
measuring at different filter voltages, the shape of the integrated energy spectrum of tritium $\beta$-decay in the 
vicinity of $E_{0}$ can be determined (for more details, see \cite{drexlin2013current}).

In order to reach the required background rate of $10^{-2}\,$cps, low-energy secondary electrons from radioactivity and 
cosmic-ray muon interactions in the vessel walls have to be shielded by magnetic and electrostatic fields that exhibit a high 
degree of axial symmetry \cite{KATRINDesignReport}. However, this technique does not shield against electrons from 
radioactive decays of neutral unstable atoms which are emanated from the inner spectrometer walls 
or from auxiliary equipment, such as pumps required to maintain the stringent ultra-high vacuum 
(UHV) conditions of $p \sim 10^{-11}\,$mbar.

Previous investigations \cite{RadonPaper, NuclearDecayPaper} have revealed that the $\alpha$-decay of 
${}^{219,220}$Rn atoms, emanated in small quantities from the installed non-evaporable getter (NEG) pump and 
from the inner surface of the stainless steel vessel, constitute a major source of background in KATRIN which cannot 
be shielded by electromagnetic fields. This background arises from electrons with energies from a few eV up to a few 
hundred keV \cite{BROWNE2001763, nancy} which accompany the primary $\alpha$-decay of both radon isotopes. 
Subsequent investigations focus on actively removing trapped high-energy electrons \cite{ECRpaper} and 
on the signature of this background by detailed simulations \cite{nancy2} and specific measurements 
\cite{RadonPaper}. Here we report on a passive background reduction technique that traps emanated radon 
atoms on a large cold surface located outside the sensitive volume of the magnetic flux tube. By confining the neutral radon atoms to 
the surface of the cold trap for an extended period of time, electrons produced by the subsequent $\alpha$-decay of 
the absorbed radon atom are magnetically shielded and cannot contribute to background processes.

Capturing radon atoms on charcoal-covered cryogenic panels is a commonly used method for 
low count rate experiments to reduce the radon-induced background rate. 
This method works well even for ${}^{222}$Rn with a half-life of 3.8\,days. 
However, in tests we found that the fine dust of charcoal released from cryo-panels can be deposited in the vicinity of the panels. With large numbers of tiny ceramic insulators inside the KATRIN spectrometers, this can lead to unwanted leakage currents.  Since we are mainly concerned about the short-lived isotopes ${}^{219,220}$Rn, our novel design employs cryo-panels made of pure OFHC copper. The mean sojourn time of a radon atom captured on the cold surface has to be long enough to allow the radon isotope to decay while still attached to the cryo-panel.    

The paper is organized as follows: First we describe the Pre-Spectrometer in its stand-alone test setup configuration 
(section~\ref{sec:Pre-Spectrometer}). Next we outline the key design criteria for a baffle to optimize the efficiency in 
adsorbing emanated radon atoms, and depict the configuration as implemented in one of the pump ports of the 
Pre-Spectrometer (section~\ref{sec:baffle}). We then report on different measurements with this improved setup, which 
demonstrate the feasibility of our method by significantly reducing the observed rate of radon-induced 
background (section~\ref{sec:baffle_measurement}). Finally, we give an outlook on how to implement this cryogenic 
technology at the much larger Main Spectrometer (section~\ref{sec:outlook}).


\section{Radon background in the Pre-Spectrometer test setup}
\label{sec:Pre-Spectrometer}

The Pre-Spectrometer (PS) is an integral component of the KATRIN beamline (see Fig.~\ref{fig:beamline}). Prior to 
its integration into the beamline, the PS was operated for over 6 years as a stand-alone test facility  
\cite{Habermehl:2009:PHD, Fraenkle:2010:PHD}, targeted at verifying and optimizing the vacuum concept 
\cite{bornschein:outgassing} and the electromagnetic design 
(\cite{PhDValerius2009}, \cite{PhDFlatt2004}) for the KATRIN spectrometers.
 
\subsection{Hardware configuration}
\label{sec:sub:technical}

\begin{figure} [t]
  \centering
  \includegraphics[width=\textwidth]{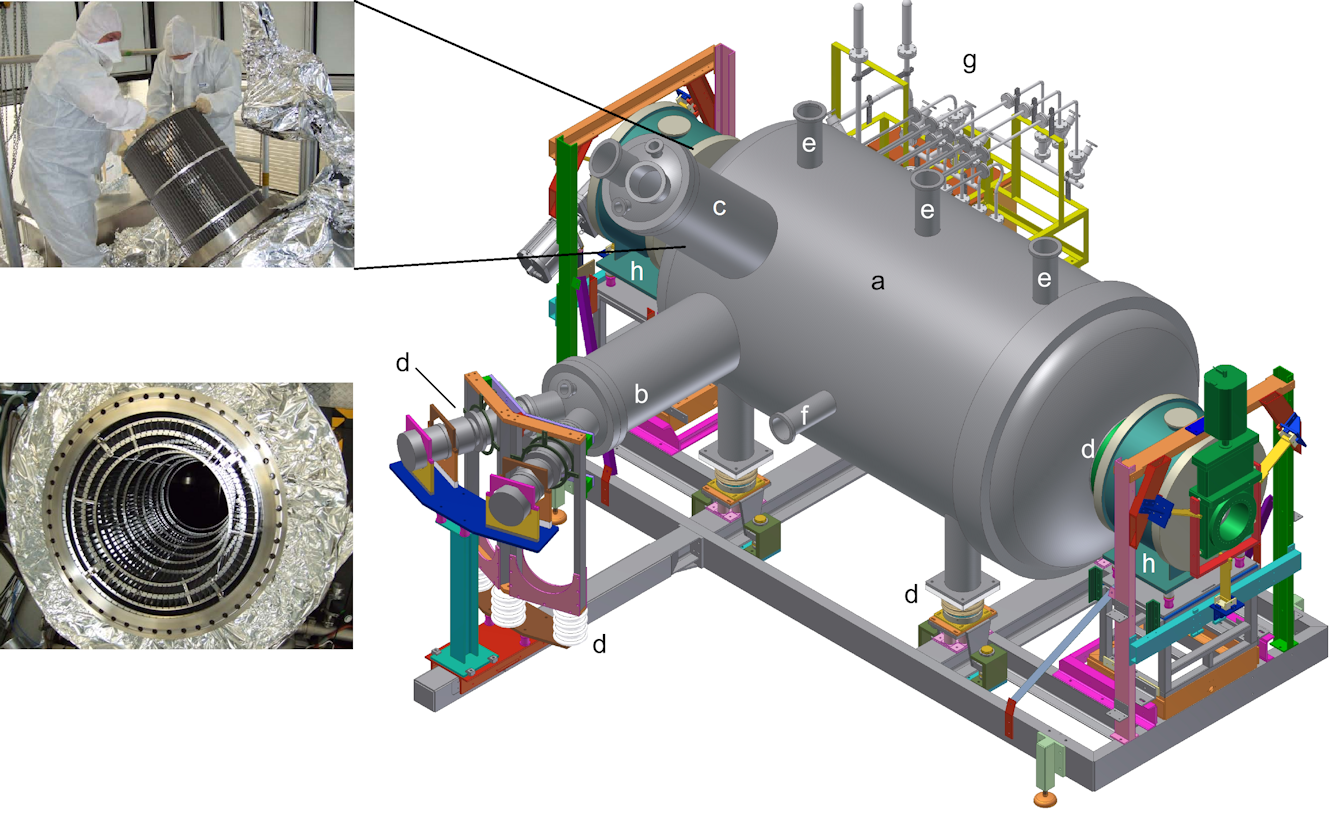}
  \caption[3D CAD model of the Pre-Spectrometer setup]{
Technical drawing of the Pre-Spectrometer setup.
    \textbf{a}) Pre-Spectrometer vessel,
    \textbf{b}) pump port at $90^\circ$ with TMPs,
    \textbf{c}) pump port at $45^\circ$ position with getter pump and baffle,
    \textbf{d}) ceramic insulators for high voltage separation,
    \textbf{e}) ports with high voltage (HV) feedthroughs for the inner electrode system,
    \textbf{f}) port for a residual gas analyzer (RGA),
    \textbf{g}) flow and return pipes for the thermo oil circuit,
    \textbf{h}) two cryogen-free superconducting solenoids.
		{The two inset photos show the non-evaporable getter (NEG) pump at the Pre-Spectrometer.
    \textbf{top:} installation into the 45$^{\circ}$-pump port.
    \textbf{bottom:} The getter cartridge in place. The $90\,$m long NEG strips are arranged in a cylindrical shape 
    \cite{Habermehl:2009:PHD}.}
		}
  \label{fig:prespec}
\end{figure}

Figure~\ref{fig:prespec} shows the cylindrical PS vessel made of $10\,\mathrm{mm}$ thick 
austenitic stainless steel (316LN). It is $3.38\,$m long with an inner diameter of 
1.68-m. Access to the inside of the spectrometer is provided by a large flange with the same 
inner diameter \cite{Habermehl:2009:PHD}. Two pump ports with a length of 
$1\,\mathrm{m}$ and a diameter of $0.5\,\mathrm{m}$ are welded to the vessel at angles of 
$45^\circ$ and $90^\circ$ with respect to the vertical axis. 
The total volume of the PS to be operated as vacuum 
chamber in the UHV pressure regime is $V_{\mathrm{PS}} = 8\,\mathrm{m}^{3}$; 
the corresponding inner surface area is $ A_{\mathrm{PS}} = 25\,\mathrm{m}^2$. The entire setup 
can be elevated to a high 
voltage (HV) of up to $-35\,$kV. The vessel acts as a Faraday cage which shields the interior from 
electric noise and interference. The inner electrodes are made of thin 
($0.2\,$mm diameter) stainless steel wires in the central part and from sheet metal (316LN) at 
both ends (see Fig.~\ref{fig:setup}).  They are operated at a slightly more negative potential 
($U_\mathrm{wire} = U_\mathrm{cone} = U_\mathrm{vessel} - 100$\,V) than the vessel, 
and allow to fine-tune the retarding potential for the integral energy analysis. 
In addition to the dominant magnetic shielding by the stray fields of the two 
super-conducting solenoids ($B_{\mathrm{max}} = 4.5\,$T) located at both ends of the PS, 
the negative offset voltage 
between inner electrodes and vessel further shields the sensitive flux tube volume from low-energy 
electrons, predominantly generated in the spectrometer walls by interactions of cosmic ray muons 
\cite{Leiber:2014:PHD}. 

For the measurements reported here, a solid-state detector was attached to the stand-alone PS test setup for signal 
read-out. The quadratic $16\,\mathrm{cm}^2$ monolithic silicon PIN diode is segmented into $64$ pixels with an area 
of $5\times 5\,\mathrm{mm}^2$ each for mapping of the sensitive magnetic flux tube 
(see Fig.~\ref{fig:measurement-warm}b) with adequate spatial resolution \cite{Wuestling2010295}. With the 
detector at ground potential and the PS at a typical retarding potential of $-18.6\,$kV, low-energy electrons generated 
in the central region of the flux tube are accelerated towards the detector, thereby gaining 
essentially the full retarding energy. Accordingly, these electrons will give rise to 
detector signals at $18.6\,$keV, well above the detection threshold of $10\,$keV 
\cite{Fraenkle:2010:PHD}.

The vacuum pumping system of the PS setup can be subdivided into two main components 
(see Fig.~\ref{fig:prespec}). First, a cascade of turbo-molecular pumps (TMPs) 
-- consisting of two TMPs (Leybold\footnote{Oerlikon Leybold Vacuum, Bonner Str. 498, 50968 
K\"oln, Germany} MAG-W 1300, nominal pumping speed for N$_2$: $1\,100\,\ell$/s) in parallel, 
followed by a 300\,$\ell$/s TMP with high compression ratio, and a scroll pump -- provides an 
effective pumping speed of $400\,\ell$/s for radon.
Second, the $45^\circ$-pump port houses a non-evaporable getter (NEG) pump with $90\,\mathrm{m}$ 
of SAES\footnote{SAES Getters S.p.A., Viale Italia 77, 20020 Lainate, Milan, Italy} St707 getter 
strips providing an effective pumping speed of about $25\,000\,\ell\mathrm{/s}$ for hydrogen \cite{benvenuti1996pumping}. With no 
moving parts, the NEG pump adsorbs hydrogen mainly by physisorption. After bake-out of the vessel at 
$200\,^\circ$C and local thermal activation of the NEG material at $350\,^\circ$C for a time period of about 
$24\,\mathrm{h}$, a pressure in the range of $\sim 10^{-10}\,\mathrm{mbar}$ has routinely been maintained in the 
PS vessel at ambient temperature \cite{bornschein:outgassing}.

\subsection{Radon emanation}
\label{sec:sub:radonemanation}

Radon emanation originates from two distinct sites in the spectrometer: the NEG material at the $45^\circ$ 
pump port, and the inner spectrometer surface, which includes auxiliary equipment such as vacuum gauges (see Fig.~\ref{fig:setup}).

The SAES St707 NEG material is a fine-grained powder made of an alloy of 
zirconium ($70\%$), vanadium ($24.6\%$), and iron ($5.4\%$) \cite{Getter}. 
The NEG powder is firmly attached to its base material, a nonmagnetic constantan alloy ($55\%$ 
copper and $45\%$ nickel) by compression bonding \cite{Getter2}. Trace amounts of progenies of ${}^{235}$U, in 
particular $^{231}$Pa and the daughters in the actinide decay chain give rise to the emanation of 
${}^{219}$Rn atoms through the porous getter material at thermal velocities, following the 
$\alpha$-decay of the mother isotope ${}^{223}$Ra. The specific ${}^{223}$Ra activity of the 
90-m long strips of the getter pump was measured to about $8\,\mathrm{Bq}$ via low-level $\gamma$-ray 
spectroscopy at MPIK in Heidelberg \cite{Fraenkle:2010:PHD}. Due to the very large inner surface of the material of 
$1\,500\,\mathrm{cm}^{2}/\mathrm{g}$ \cite{Getter2} and its fine-grained structure, the emanation  of 
radon atoms into the PS is very efficient, resulting in a measured released ${}^{219}$Rn activity of 
about $\sim 8\,\mathrm{mBq}$ (corresponding to an emanation probability of $0.1\%$) \cite{RadonPaper}. Additional 
radon emanation occurs from the large electro-polished inner surface of the Pre-Spectrometer vessel and in particular 
from its weld seams, which can contain trace amounts of the actinides in the ${}^{235}$U, ${}^{232}$Th, and ${}^{238}$U series, giving rise to highly volatile ${}^{219}$Rn-, ${}^{220}$Rn- and 
${}^{222}$Rn-atoms, respectively. In \cite{nancy2} the different specific activities inside the vacuum vessel 
of ${}^{219}$Rn from the NEG material and the vessel's walls, as well as ${}^{220}$Rn from the walls were 
determined to be $(0.88\pm0.22)\,\mathrm{mBq/m}^3$, $(0.26\pm0.26)\,\mathrm{mBq/m}^3$, and 
$(6.53\pm2.18)\,\mathrm{mBq/m}^3$, respectively (see also Tab.~\ref{tab:radon}), when no 
vacuum pumps are connected to the vessel. Switching on the pumps will reduce the specific 
activity, since some of the radon isotopes will be pumped out before they decay. These 
quantitative results agree with earlier measurements in \cite{RadonPaper} and show the dominant 
radon emanation channel to be that of ${}^{220}$Rn from the bulk material of the vessel, followed 
by ${}^{219}$Rn from the NEG material.

During the standard operation of the PS emanated radon atoms will either decay inside the PS 
(see Tab.~\ref{tab:radon}) or they will be pumped out before undergoing $\alpha$-decay. 
Several half-lives $t_{1/2}(\rm Rn)$ after switching on the vacuum pumps, the activity 
$A_{\rm Rn} =\lambda_{\rm Rn}\cdot N_{\rm Rn}^{\rm PS}$ inside the PS approaches a constant value that
depends on the decay rate $\lambda_{\rm Rn} = \ln (2)/t_{1/2}(\rm Rn)$ and the average number of
 radon atoms  $N_{\rm Rn}^{\rm PS}$ in the main volume $V_{\rm PS} = 8$\,m$^3$.  
For this equilibrium state between emanation, decay and pump-out, the decay probability 
$P_{\rm Rn}^{\rm PS}$, defined as the ratio between the activity and the emanation rate 
$Q_{\rm Rn}^{\rm PS}$ into the PS, can be estimated by the ansatz:

\begin{eqnarray}
\frac{{\rm d}N_{\rm Rn}^{\rm PS}}{{\rm d}t} & = & Q_{\rm Rn}^{\rm PS} - 
         \lambda_{\rm Rn}\cdot N_{\rm Rn}^{\rm PS} - 
         \frac{S_{\rm Rn}}{V_{\rm PS}}\cdot N_{\rm Rn}^{\rm PS} = 0 \label{Eqn:DGL} \\
P_{\rm Rn}^{\rm PS} & = & \frac{A_{\rm Rn}}{Q_{\rm Rn}^{\rm PS}} =
          \frac{\lambda_{\rm Rn}\cdot N_{\rm Rn}^{\rm PS}}{Q_{\rm Rn}^{\rm PS}} =
          \frac{\lambda_{\rm Rn}}{\lambda_{\rm Rn} + S_{\rm Rn}/V_{\rm PS}}. \label{Eqn:decay_probability}
\end{eqnarray}

The total effective pumping speed $S_{\rm Rn}$ is the sum of all effective pumping speeds attached to the volume, 
including the conductance losses into neighboring components, such as the Main Spectrometer and the CPS in the 
final configuration.

Due to their different half-lives, the decay probabilities of 
${}^{219}$Rn and ${}^{220}$Rn, listed in Tab.~\ref{tab:radon}, differ by a factor of 
$\sim 4$. The isotope ${}^{222}$Rn is pumped out of the spectrometer by the TMPs well before it can decay 
and hence can be neglected. Its short half-life and the resulting large decay probability contribute 
to the fact that ${}^{219}$Rn decays dominate the background rate, although the total emanation rate of 
${}^{219}$Rn is much lower than the one of ${}^{220}$Rn (see Tab.~\ref{tab:radon}). 
This dominant role of ${}^{219}$Rn in generating background can be explained in the 
framework of our radon background model described in \cite{nancy}, and briefly summarized in the following subsection.

\begin{table}[t]
\footnotesize
\centering
\caption[radon half-lifes]{Half-lifes of different radon isotopes and measured \cite{nancy2} radon emanation rates for 
${}^{219}$Rn and ${}^{220}$Rn from the NEG material and the inner surface of spectrometer. The decay 
probability for each isotope before pump-out by two TMPs is given for the effective pump speed 
$S_{\mathrm{Rn}}=400\,\ell\mathrm{/s}$ and the total Pre-Spectrometer volume 
$V_{\mathrm{PS}} = 8\,\mathrm{m}^3$. The background rates of ${}^{219,220}$Rn have been scaled to 
$V_{\mathrm{PS}}$ from the observed background rate at the detector, taking into account the fraction of the 
surveyed volume ($7\%$) and assuming a homogeneous decay distribution of radon \cite{nancy2}. The background 
rates correspond to the average background rates observed by Fr\"ankle et.al. \cite{RadonPaper} with one TMP in 
operation. -- check numbers for 8\,m$^3$ --}
\label{tab:radon}
\begin{tabular}{l l l l}
\hline
& ${}^{219}$Rn & ${}^{220}$Rn & ${}^{222}$Rn \\
\hline
half-life (s)	 & $3.96$ & $55.6$ & $3.3\cdot 10^{5}$ \\
total emanation NEG (mBq)  & $7.5 \pm 1.8$ & - & - \\
total emanation bulk 	(mBq)  & $2.2 \pm 2.2$ & $55.5 \pm 18.5$ & - \\
decay probability (2 TMPs) & $0.79$ & $0.21$ & $4.4\cdot 10^{-5}$ \\
background rate ($10^{-3}\,$cps) & $25 \pm 8$ & $2.1\pm0.4$ & - \\ 
\hline
\end{tabular}
\end{table}

\subsection{Radon background}
\label{sec:sub:radonbackground}

During the $\alpha$-decay process of a radon isotope the primary $\alpha$-particle is accompanied by X-ray photons 
and electrons ejected from the atomic shells with energies ranging from a few eV up to several hundred keV 
\cite{nancy, Wandkowsky:2013:PHD}. It is important to note that the $\alpha$-particle, the positively charged 
daughter ion (${}^{215,216}$Po) and fluorescence X-rays after $\alpha$-decay are no 
background concerns, as they immediately hit the inner walls of the vessel or the inner 
electrodes. Secondary electrons, generated there upon impact are strongly suppressed by the excellent magnetic and 
electrostatic shielding of the spectrometer. The backgrounds relevant for a MAC-E filter are associated with 
extremely fast electron emissions ($<10^{-13}\,$s) accompanying the $\alpha$-decay. These result from a 
variety of nuclear and atomic de-excitation processes such as internal conversion or shake-off, and the slower restructuring of the 
disturbed atomic shell \cite{nancy}. If, for instance, a vacancy is created in an inner shell (either by a shake-off 
process or internal conversion), it will subsequently be filled with electrons from a higher atomic shell. The corresponding 
energy is then emitted either as an X-ray fluorescence photon or an Auger (or Coster-Kronig) electron with typical 
energies of up to several keV. These processes are identical for both radon isotopes. However, the probability for 
emission of very high-energy electrons is larger for ${}^{219}$Rn than for ${}^{220}$Rn; for example, the total 
probability for internal conversion of the ${}^{219}$Rn daughter ${}^{215}$Po${}^{\ast}$ is about $3\%$, releasing 
mainly electrons with energies of $178\,\mathrm{keV}$ $(1.27\%)$ and $254\,\mathrm{keV}$ 
$(0.74\%)$ \cite{BROWNE2001763}.

\begin{figure}[t]
  \centering
  \includegraphics[width=15cm]{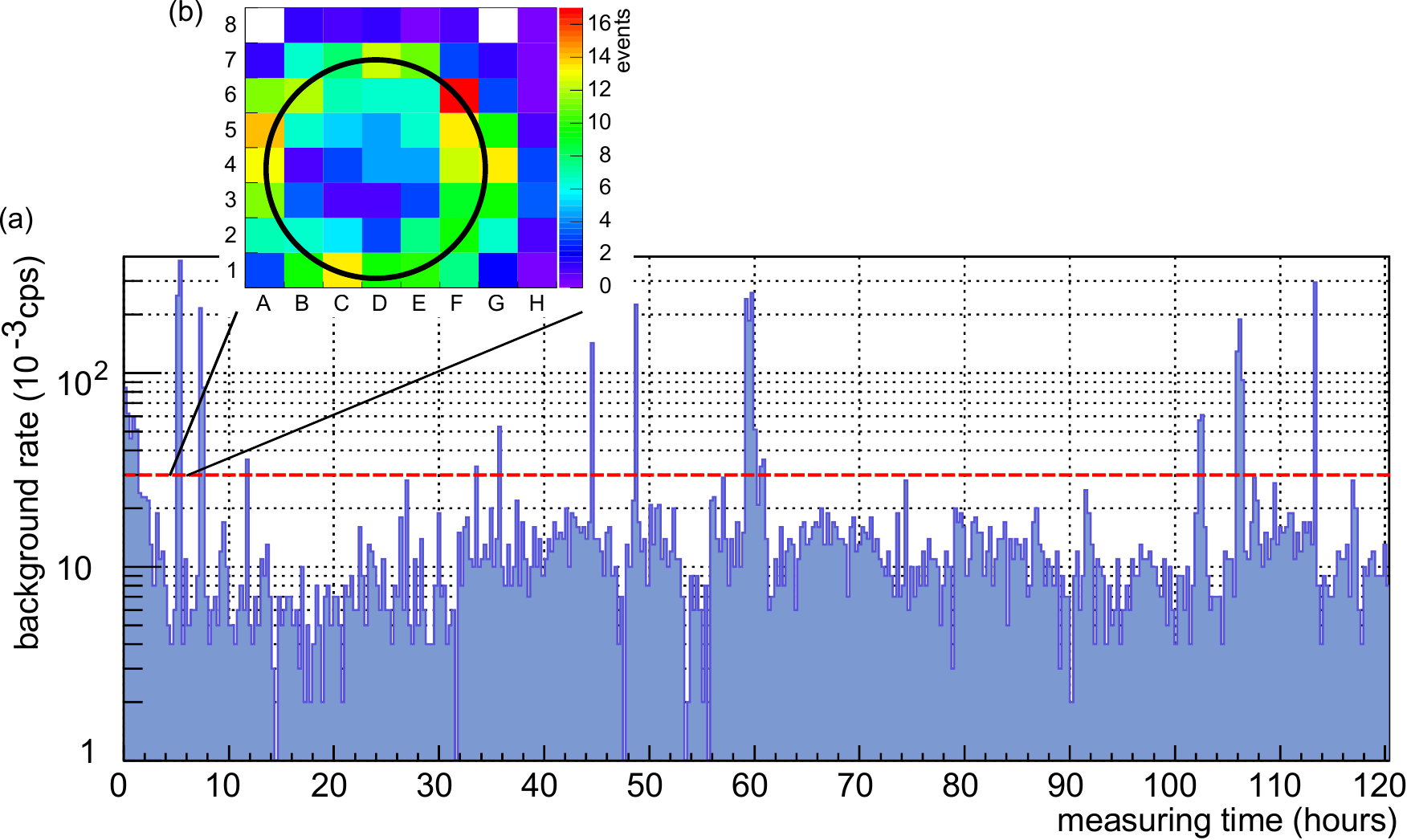}
  \caption[measurement with warm baffle]{Measurement $1$: \textbf{(a)} observed background rates in $1000\,$s time 
  intervals with the baffle operating at room temperature. The rates include the intrinsic detector background 
  of $(6.3 \pm 0.2) \cdot 10^{-3}\,\mathrm{cps}$. The dashed line at $0.03\,$cps discriminates radon spikes from 
  Poisson-distributed background (see Fig.~\ref{fig:baffle-analyse})\ \textbf{(b)} The pixel distribution at the 
  detector of a radon spike started $4.95\,$h into the run, persisted for $33$ minutes, and showed a ring-like 
  topology with 750 detector hits. The superimposed ring fit (black circle) demonstrates that the pattern originates 
  from the magnetron drift of a trapped high-energy electron. At the time of the measurement, the detector was 
  positioned off-center, leading to the apparent lateral shift of the ring fit around the central axis of the spectrometer.}
    \label{fig:measurement-warm}
\end{figure}

An electron in the high-energy range above few keV has a very high probability to be trapped inside the 
spectrometer by the magnetic mirror effect (see Fig.~\ref{fig:setup}) inherent to a MAC-E filter 
\cite{higaki2008electrons}. Accordingly, a single trapped multi-keV electron can generate up to several hundreds of 
low-energy secondary electrons via residual gas ionization. These secondary electrons can leave the magnetic bottle on 
rather short time-scales (of the order of a few seconds \cite{Wandkowsky:2013:PHD} for the pressure regime 
obtained here). If released in downstream direction, they reach the detector with an energy close 
to the retarding potential, thus contributing to the 
background rate in the region-of-interest. This results in an elevated level of background over an extended period of 
time\footnote{lasting from minutes to hours, depending on the primary energy and on the residual gas pressure 
(see \cite{nancy2}).}, which is terminated only if the storage condition of the primary electron is broken as a result of 
a scattering process or non-adiabatic motion, or if the electron has cooled down to a value below the ionization 
threshold. This temporary increase of the background rate can be detected with the $8\times8$ PIN diode array, 
resulting in a so-called `radon spike' (see Fig.~\ref{fig:measurement-warm}) that has a characteristic ring-like event 
topology due to the magnetron drift of the stored electron, as visible in 
Fig.~\ref{fig:measurement-warm}(b). 
Detailed information on this observed pattern can be found in 
\cite{RadonPaper, NuclearDecayPaper, nancy2}. However, in the majority of 
$\alpha$-decay processes, the atomic 
shells are left largely unperturbed; an average of two low-energy electrons are emitted which share an excitation 
energy of about $250\,\mathrm{eV}$ \cite{nancy}. These electrons do not produce a spike and 
contribute to the so-called `singles' background characterized by a Poisson-distributed rate.

Radon-induced backgrounds in the PS and the MS will limit the neutrino mass 
sensitivity of KATRIN, if no countermeasures are taken. 
Apart from methods that actively remove stored electrons, 
the passive technique of adsorbing emanated radon atoms onto a cold surface outside the active 
flux tube (see Fig.~\ref{fig:PS_Baffle}) is a very efficient method to reduce this 
background, as will be discussed in the following section. 


\section{Design and efficiency of a liquid nitrogen cooled baffle}
\label{sec:baffle}

In this section we first outline the general design criteria for a cold trap operating
in an electrostatic spectrometer (section~\ref{sec:sub:criteria}), before we describe the 
baffle implemented in the Pre-Spectrometer test set-up (section~\ref{sec:sub:baffle-setup}).
Finally we calculate the efficiency of removing emanated radon atoms by 
cryo-sorption in a cryogenic baffle, mounted in front of the NEG pump inside the 
45$^\circ$ pump port (section~\ref{sec:sub:efficiency}).

\subsection{Baffle design criteria}
\label{sec:sub:criteria}

The efficient removal of ${}^{219,220}$Rn atoms from the spectrometer volume requires an optimized design 
of the baffle. In the following, we outline key design criteria for the baffle geometry, surface material and shape, 
as well as the cryogenic operation:

\begin{itemize}
	\item {\bf Position and shape:} the position and the shape of the baffle have to be such that the direct line of 
sight from the getter strips of the NEG pump to the sensitive flux tube volume of the spectrometer is blocked (see 
Fig.~\ref{fig:prespec} and Fig.~\ref{fig:PS_Baffle}). In this case, ${}^{219}$Rn atoms emanating from the NEG 
pumps have to impinge at least once on the cold surface of the baffle, where they are adsorbed before entering 
the inner vessel volume. For radon emanating from the inner 
vessel surface, the position of the baffle is not crucial, as radon atoms are emanating with thermal velocities 
and thus follow stochastic trajectories before hitting the baffle surface.
	
\item {\bf H${}_{2}$ pumping:} hydrogen, predominantly from outgassing of the vessel walls, 
constitutes the dominant gas load that limits the ultimate pressure in the vessel. 
While fulfilling the above design criterion for radon atoms, the baffle has to 
guarantee a high conductance for molecular hydrogen to reach the NEG pump for effective 
pumping. This is of crucial importance to maintain excellent UHV conditions in the spectrometer.

	\item {\bf Surface size:} the mean time interval after emanation until a radon atom hits the cold baffle should 
be smaller than its half-life. At room temperature, radon atoms move in vacuum with an average speed  
$\bar{c} = 168$\,m/s. With a path length of 2\,--\,3\,m between two successive hits of the wall, an atom hits the 
wall about 70 times per second. Therefore a coverage of the cold surface of about 1\% would be sufficient for the 
longer lived $^{220}$Rn, which is the dominant isotope emanated from the vessel walls. The baffle in the opening of 
the pump port has a diameter of 50\,cm, which corresponds to a coverage of 0.8\% of the 25\,m$^2$ surface of 
the PS. 

	\item {\bf UHV cleanliness:} the UHV conditions require stringent cleanliness protocols for all materials  
installed inside the KATRIN spectrometers \cite{WolfVacuum}. The baffle has to withstand bake-out temperatures 
of up to $350\,^\circ$C. The material itself should also 
display only moderate rates of hydrogen outgassing and virtually no radon emanation.
	
	\item {\bf Emissivity and heat conduction:} the material should have a low emissivity $\eta < 0.05$ to maintain a 
uniform operating temperature $T_{\mathrm{op}}$ over the entire baffle surface, despite the considerable heat load 
from the large spectrometer surface at room temperature. By the same token, the material 
should be an excellent heat conductor to maximize the heat transfer from the cooling fluid to the adsorbing surface.

	\item {\bf Adsorption enthalpy:} the mean sojourn time $\tau_{\mathrm{des}}$ of ${}^{219}$Rn and ${}^{220}$Rn atoms
is an important property of a cryogenic baffle. It is defined by the adsorption enthalpy of the baffle material 
$\Delta H_{\mathrm{ads}}$, the maximum phonon frequency $\nu_{\mathrm{b}}$, and the temperature $T_{\mathrm{baf}}$ 
of the baffle surface \cite{RadonAdsorption}:
\begin{equation}
	\tau_{\mathrm{des}} = \frac{1}{\nu_{\mathrm{b}}} \mathrm{exp}
                \left(\frac{-\Delta H_{\mathrm{ads}}}{R\cdot T_{\mathrm{baf}}}\right) \label{eqn:tau_des}
\end{equation}
with $R$ denoting the gas constant ($8.314\,\mathrm{J}\,\mathrm{K}^{-1}\mathrm{mol}^{-1}$). 
The baffle material should possess a large 
$\Delta H_{\mathrm{ads}}$, resulting in a long sojourn time; the longer the sojourn time, the higher the 
probability of a radon atom decaying while still adsorbed on the cold baffle.
The adsorption enthalpy of a clean surface can be reduced over time by various processes, 
such as oxidization and deposition of contaminants (e.g. water), even in a UHV environment. 
In our specific case, exposure of the chosen baffle material to ambient air during venting of the spectrometer should not 
reduce $\Delta H_{\mathrm{ads}}$ below the required limit. Finally, $\Delta H_{\mathrm{ads}}$ should be large enough 
so that liquid nitrogen (LN$_2$) can be used as cooling fluid to obtain a sufficiently large mean sojourn time 
$\tau_{\mathrm{des}}$ (in particular for longer-lived ${}^{220}$Rn isotopes). 

	\item {\bf Cooling fluid:} the supply lines for the cooling fluid have to be designed for operation in a UHV 
environment in the presence of high voltage, so that the baffle (together with the spectrometer vessel) can be 
operated at a high potential of up to $-35\,$kV, while the large volume LN$_2$ reservoir is at ground potential.
\end{itemize}  

\noindent To first order, a baffle made of cleaned OFHC copper located at the entrance of the NEG pump port 
(see Fig.~\ref{fig:PS_Baffle}a) appears to be a viable solution, as this material fulfills all aforementioned requirements.

\subsection{Baffle setup}
\label{sec:sub:baffle-setup}

\begin{figure}[t]
  \centering
  \includegraphics[width=15cm]{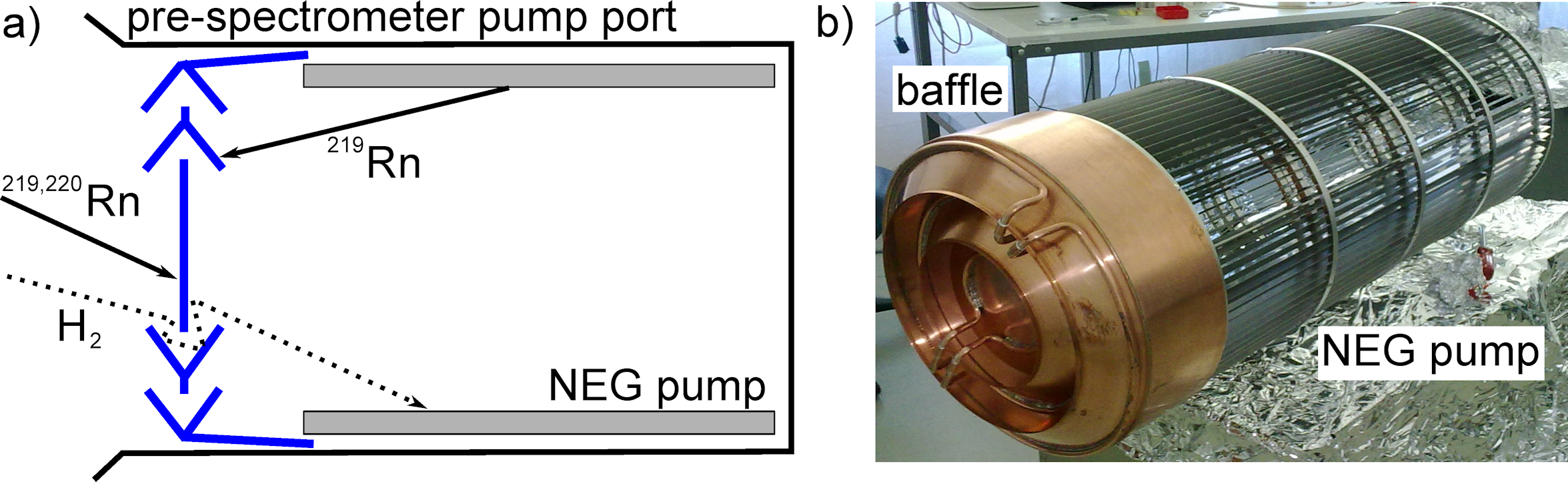}
  \caption[Pre-Spectrometer Baffle]{The Pre-Spectrometer baffle setup: \textbf{(a)} schematic drawing of the cross-section 
of the PS pump port with the NEG pump (length: 100\,cm, diameter: 50\,cm) and liquid-nitrogen-cooled baffle. \textbf{(b)} The 
baffle is attached mechanically to the NEG pump.}
  \label{fig:PS_Baffle}
\end{figure}

Based on the general design criteria and efficiency estimates listed above, a baffle prototype using untreated polished OFHC 
copper as adsorbing surface was installed in the PS test setup in front of the NEG pump, as shown in 
Fig.~\ref{fig:PS_Baffle}(a). The copper baffle covers an effective circular surface area of $0.2\,\mathrm{m}^2$, thus 
covering 0.8\% of the surface of the PS. It was designed and manufactured by the company HSR\footnote{HSR AG, 
Foehrenweg 16, P.O. Box 109, 9496 Balzers, Principality of Liechtenstein}. With a diameter of $478\,\mathrm{mm}$ and a 
length of $162\,\mathrm{mm}$, the baffle is connected at four points directly to the NEG pump and thus can be installed as 
a single unit into the $45^\circ$ PS pump port (see Fig.~\ref{fig:PS_Baffle}b). The central design elements consist of two 
circular V-shaped folds ($112\,$mm high and $59\,$mm deep) with a surface area of $0.7\,\mathrm{m}^2$ and a central 
disk with a diameter of $236\,$mm (area $0.09\,\mathrm{m}^2$), which together block all direct lines-of-sight from the 
surfaces of the NEG pump to the inner surface of the PS. Both elements are manufactured from $1\,$mm thick copper 
sheets with an overall mass of about $7\,$kg and a total surface area of $\sim 0.8\,\mathrm{m}^2$. In order to maintain 
the cold trap below $90\,$K, a LN$_2$ flow of about $10\,\ell\mathrm{/h}$ is required. For 
continuous cryogenic supply to the baffle, a UHV feedthrough for liquid nitrogen is used. To compensate for thermal 
expansions, a two-loop spiral is integrated in both supply and return lines inside the pump port.

The operation of the PS as a MAC-E-filter requires the whole setup, including the baffle and its supply line, to be elevated to 
a negative potential of more than $-18\,\mathrm{kV}$. Since the electrically insulating LN$_2$ connections were not available,
a $200\,\ell$ liquid nitrogen supply dewar, temporarily used in the 
test measurements, was mounted on four $30\,\mathrm{cm}$-high ceramic insulators and set to the same potential as the 
vessel. The temperature of the baffle was continuously monitored at the exhaust pipe by a PT100 sensor to ensure that the 
radon adsorption probability remains constant over time.

As outlined above, the operation of the baffle comes at the expense of a reduced 
effective pumping speed of the NEG pump for molecular hydrogen, the dominant residual gas species. For this baffle design, 
the reduced conductance of the pump port should result in an increase of the final equilibrium pressure inside the 
Pre-Spectrometer by a factor of 3.6\footnote{Based on a TPMC simulation by Xueli Luo with the software 
package \textit{ProVac3D} \cite{ProVac3D}, developed at KIT (ITeP).}.  Retaining a high pumping 
speed for hydrogen is of particular importance for neutrino mass measurements with the Main Spectrometer. 
However, an optimized hydrogen pumping efficiency is less important for the 
measurements discussed below. This is due to the fact that the efficiency for radon trapping by adsorption on the cold 
copper baffle surface is largely independent of the absolute pressure, as long as UHV conditions ($\sim 10^{-9}\,$mbar) are 
maintained to suppress rapid formation of surface adsorbents, such as by H${}_{2}$O. These surface adsorbents 
would lower the adsorption 
enthalpy of a clean oxidized copper surface. In all our measurements, a free molecular flow of residual gas atoms and 
molecules was maintained, as was assumed in the simulations below.

\subsection{Baffle efficiency}
\label{sec:sub:efficiency}

\begin{figure}[t]
  \centering
  \includegraphics[width=11.5cm]{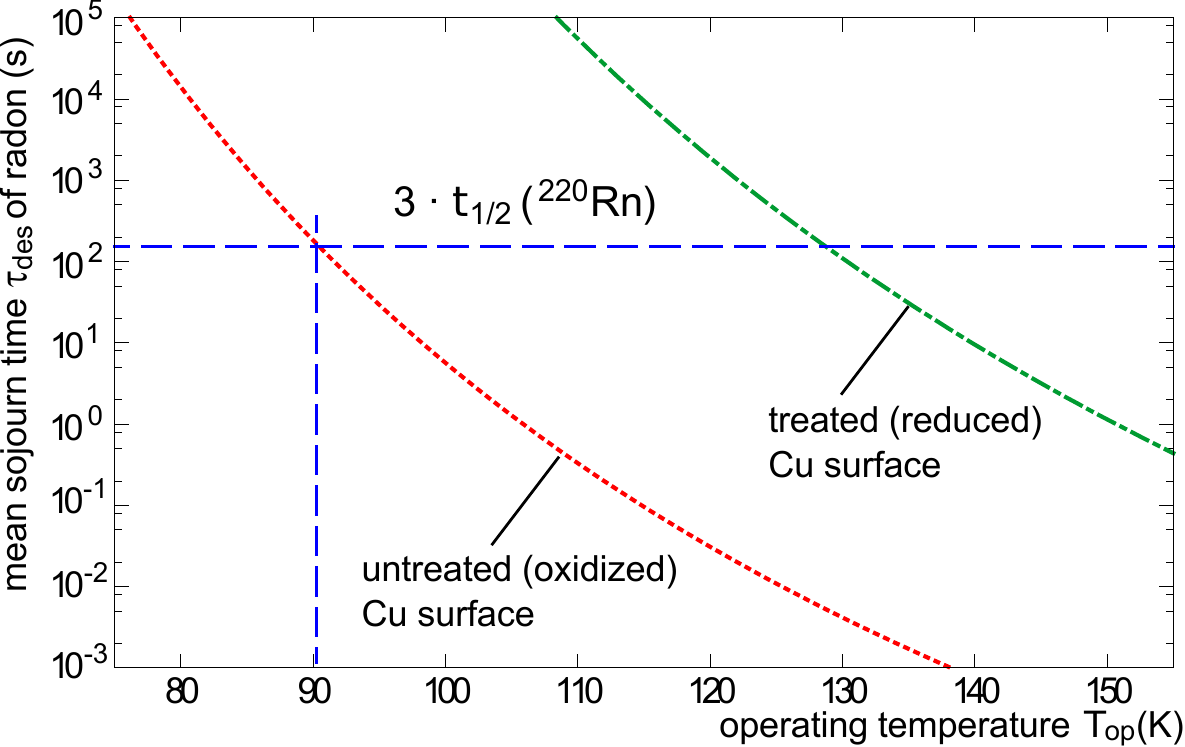}
  \caption[Sojourn time]{Mean sojourn time $\tau_{\mathrm{des}}$ of single ${}^{219,220}$Rn atoms on different 
surface-treated pure copper sheets as a function of the baffle operating temperature $T_{\mathrm{op}}$. The dotted line 
applies to the case of untreated oxidized copper surfaces with an adsorption enthalpy  
$\Delta H_{\mathrm{ads}} = -26\,$kJ/mol. This is our baffle material. The chain dotted line pertains 
to copper surfaces heated for $2\,$h at $1000\,$K in a ($90\%:10\%$) N${}_{2}$+H${}_{2}$ atmosphere, 
corresponding to $\Delta H_{\mathrm{ads}} = -37\,$kJ/mol. Experimental data for the enthalpies are taken from 
\cite{RadonAdsorption}. The horizontal line corresponds to our benchmark for the sojourn time 
$\tau_{\mathrm{des}} \geq 3\cdot t_{1/2}$ (${}^{220}$Rn). The intersection of the curve for untreated copper 
with this line defines the upper range of baffle operating temperatures $T_{\mathrm{op}}<90\,$K.}
    \label{fig:temperature}
\end{figure}

When estimating the efficiency of radon suppression in the PS, we have to consider two cases: (i) the reduction of radon 
atoms inside the main volume of the PS, and (ii) the transmission of radon atoms from the NEG pump behind the baffle 
into the main volume.

As pointed out before, the sojourn time $\tau_{\mathrm{des}}$ (Eqn.~\ref{eqn:tau_des}) for radon atoms is the 
single most important property of a cryogenic baffle. With the temperature fixed to the LN$_2$ temperature, 
and $\nu_{\mathrm{b}} \approx 10^{-13}$\,Hz for most metals, $\Delta H_{\mathrm{ads}}$ defines the 
efficiency of a cryogenic baffle.  Extensive measurements of $\Delta H_{\mathrm{ads}}$ by observing radon adsorption on 
non-treated and treated copper, as well as other metal surfaces, have been performed by Eichler et al. 
\cite{RadonAdsorption}. In the following, we make use of the experimental data and discussions of this work. 
Since the baffle will be exposed to air during installation, we assume a thin oxide layer on the surface, 
independent of the cleaning procedure after manufacturing. 
In Tab.~1 of \cite{RadonAdsorption} experimental and theoretical data for $\Delta H_{\mathrm{ads}}$ 
for 6 different untreated metal surfaces are compared. The case of an untreated (oxidized) copper 
surface is found to be superior to other untreated metal surfaces by exhibiting the largest  
$\Delta H_{\mathrm{ads}}(\mathrm{therm}) = -26\,$kJ/mol (see Fig.~\ref{fig:temperature}). 

The ratio between the surface areas of the baffle and the spectrometer vessel determines the probability of radon atoms to 
undergo $\alpha$-decay before being adsorbed on the baffle. After emanation from the vessel walls, a radon atom travels 
with a mean thermal velocity $\bar{c} =168$\,m/s in a stochastic motion and will impinge eventually onto the cold surface 
of the baffle after some time $\tau_{\mathrm{hit}}$. This time can be estimated from $V_{\mathrm{PS}}$ and the flow 
rate

\begin{equation}
 Q_\mathrm{pp} = \frac{1}{4}\,\bar{c} \cdot A_{\rm pp} = 8247\,\ell/\mathrm{s} \label{equ:Qpp}
\end{equation}
 
\noindent into the opening of the pump port $A_{\rm pp} = 0.196$\,m$^2$, housing the baffle: $\tau_{\mathrm{hit}} = V_{\mathrm{PS}}/8247\,\ell/{\rm s} \cong 1\,$s.

For ${}^{220}$Rn atoms emanating from the vessel surface this short time span implies a rather small 
decay probability $P_{\rm Rn}^{\rm PS} = 1.2\%$ before hitting the 
baffle. Due to the much shorter half-life of ${}^{219}$Rn, its probability of decaying within the first 
second is 16\%. Thus, the six times larger emanation rate of ${}^{220}$Rn atoms in the PS (see Tab.~\ref{tab:radon})
is counterbalanced by the 13 times smaller $P_{\rm Rn}^{\rm PS}$ compared to ${}^{219}$Rn. 

However, a radon atom has only a finite probability to stick to the baffle after hitting it.
For a more accurate estimate of the effect of a cold baffle on $P_{\rm Rn}^{\rm PS}$ (Eqn.~\ref{Eqn:decay_probability}), 
the effective pumping speed $S_{\rm Rn}$ of the baffle has to be calculated. In the molecular 
flow regime, the effective pumping speed of a vacuum pump can be described in accordance with Eqn.\ref{equ:Qpp}, 
as $S_{\rm eff} = Q_\mathrm{pp} \cdot w$,
with the pumping probability $w$ for a particle flying through the opening of the pump port.  

Since the baffle in the pump port is flush with the inner wall of the vessel, and the baffle panels cover the whole 
cross section of the pump port, a rough approximation of the pumping probability $w$ would be the effective 
sticking coefficient  $\alpha_{\rm eff}$. It quantifies the fraction of radon atoms hitting the surface, which stick to it 
and decay while still attached to the surface of the baffle. The decay of a radon atom on the baffle's surface, 
outside of the magnetic flux tube of the PS, is equivalent to being pumped out. The effective sticking coefficient 
$\alpha_{\rm eff}$ has been introduced and defined in \cite{MSradonsimu2017}. It depends only on the ratio 
of the sojourn time $\tau_{\rm des}$ and the mean lifetime of the radon isotope 
$\tau_{\mathrm{Rn}} = t_{1/2}(Rn)/\ln(2)$: 

\begin{equation}
\alpha_{\rm eff} = \alpha_0\cdot \frac{1}{1 + \tau_{\rm Rn}/\tau_{\rm des}}.
         \label{equ:alpha_eff}
\end{equation}

\noindent with the nominal sticking coefficient $\alpha_0 \approx 1$, which is defined as the probability that a radon atom 
hitting the cold surface of the baffle is adsorbed. 

A more accurate approach for determining the pumping
probability $w$ is a Test Particle Monte Carlo (TPMC) simulation (MolFlow+ \cite{Molflow2016}) of the geometry of 
the baffle (Fig.~\ref{fig:PS_Baffle}), which uses $\alpha_{\rm eff}$ as a property of the surface elements in the model. In general, 
the simulated value of $w$ is larger than $\alpha_{\rm eff}$, since there is always a chance that the re-desorbed 
particle hits another section of the cold baffle, instead of escaping into the main volume of the PS. 
For larger $\alpha_{\rm eff}$ both values converge, since most radon atoms have decayed already after the first hit.

\begin{figure}[t]
{  \centering
  \includegraphics[width=\textwidth]{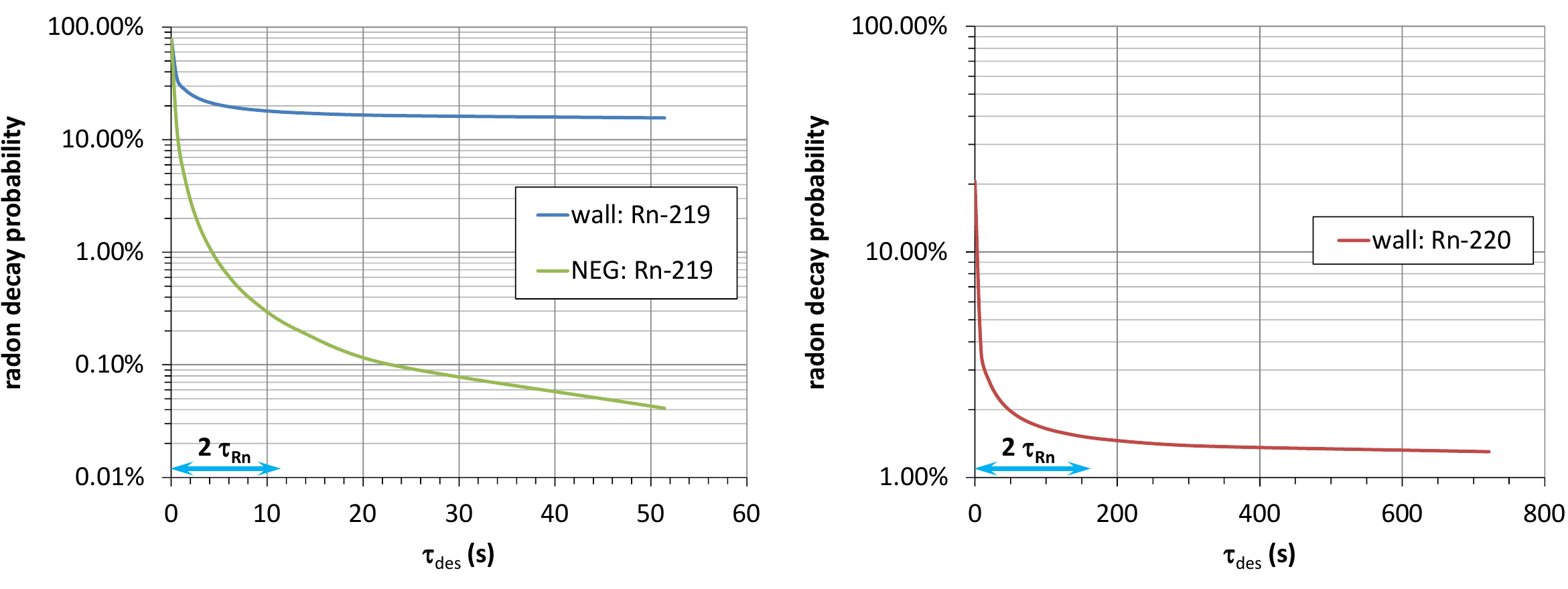}}
  \caption[Radon decay probability]{Simulated radon decay probability for $^{219}$Rn (left plot) and $^{220}$Rn. 
(right plot). For $\tau_{\mathrm{des}} \ge 2\cdot \tau_{\mathrm{Rn}}$ the radon reduction efficiency quickly reaches
its limits for radon emanating from the walls. For $^{219}$Rn from the NEG pumps the decay probability is 
mainly reduced by the low transmission probability through the cold baffle. At the sojourn time of  
$2\cdot \tau_{\mathrm{Rn}}$ it is already reduced by a factor of 300. }
    \label{Fig:baffle_reduction}
\end{figure}
   
Fig.~\ref{Fig:baffle_reduction} shows the dependence of $P_{\rm Rn}^{\rm PS}$ over the sojourn time 
$\tau_{\mathrm{des}}$ for $^{220}$Rn and $^{219}$Rn emanating from the wall, and for $^{219}$Rn from the 
NEG pumps. For a warm baffle only the TMPs pump out radon; the decay probability converges to 
the numbers given in Tab.~\ref{tab:radon}. For lower temperatures where $\tau_{\mathrm{des}}$ exceeds 
$2\cdot \tau_{\mathrm{Rn}} \approx 3\cdot t_{1/2}({\rm Rn})$ ($\alpha_{\rm eff} \ge 67\%$), the 
decay probability for radon inside the main volume is already close to the minimum values of 1.24\% for $^{220}$Rn 
and 14.9\% for $^{219}$Rn ($\alpha_{\rm eff} = 100\%$). This leads to a benchmark value of 160\,s for the sojourn 
time of $^{220}$Rn that should at least be reached with the material of the baffle at liquid nitrogen temperature.  
For $\tau_{\mathrm{des}}$ above the benchmark time, the reduction efficiency is mainly limited by the ratio 
between the surfaces of the baffle and the PS vessel. 
Below the benchmark time the length of the sojourn time starts to dominate the reduction efficiency. 

Radon atoms emanating from the NEG pump have to pass the baffle before 
entering the main volume. Since most radon atoms hit the baffle surfaces more than once before reaching the 
main volume, the transmission probability is much lower than the pumping probability. Above a sojourn time of  
$\tau_{\mathrm{des}} = 2\,\tau_{\mathrm{Rn}} = 11.4$\,s, the decay probability for $^{219}$Rn is already 
reduced by a factor of 300, as shown in the left plot of Fig.~\ref{Fig:baffle_reduction}. 
The decay probability for Radon emanating from the walls inside the main volume is reduced by a factor of 4.4. 
With the $^{219}$Rn rates given in Tab.~\ref{tab:radon} the background rate for this moderate value of $\tau_{\mathrm{des}}$ is
$1.4\cdot 10^{-3}$\,cps, completely dominated by radon emanating from the walls. For $^{220}$Rn from the walls 
this small value of $\tau_{\mathrm{des}}$ already reduces the decay probability inside the main volume 
from 20.6\% for warm baffles to 3.5\% (see Fig.~\ref{Fig:baffle_reduction}). This reduces the contribution of 
$^{220}$Rn to the total 
background rate to $4\cdot 10^{-4}$\,cps. This first estimate, based on simulations, shows already that 
even a moderate sojourn time of radon on the baffle would sufficiently reduce the radon-induced 
background rate in the Pre-Spectrometer.  


\section{Measurements with the baffle system}
\label{sec:baffle_measurement}

After successful pressure and leak testing of the baffle, it was attached to the frame of the NEG pump and mounted in the PS pump port. Three long-term measurements (see Tab.~\ref{tab:PS_Baffle}) were performed, targeted to validate the suppression of radon-induced background by a liquid-nitrogen-cooled baffle. All measurements were performed using identical PS operating conditions: a pressure regime of $\sim 3 \cdot 10^{-9}\,\mathrm{mbar}$\footnote{The NEG pump was not activated after installing the baffle. The increased equilibrium pressure facilitated the observation of the signature `rings' of the radon spikes (Fig.~\ref{fig:measurement-warm}) by minimizing the cooling time of the stored KeV-energy electrons. The corresponding pressure level is close to the value of the measurements in \cite{nancy2}, labeled `HPG' for high pressure with getter strips installed.}; a symmetric magnetic field setup with both superconducting solenoids operating at $B = 4.5\,\mathrm{T}$, resulting in a value of $B_{\mathrm{min}} = 0.016\,$T \cite{Habermehl:2009:PHD} in the central analyzing plane; and an electric potential of the vessel $U_{\mathrm{vessel}} = -18.0\,\mathrm{kV}$. The inner electrodes were elevated on a slightly more negative potential $U_{\mathrm{electrodes}} = -18.5\,\mathrm{kV}$ to fine-tune the electrostatic retarding potential and to eliminate muon-induced background electrons from the vessel walls \cite{Fraenkle:2010:PHD}. The region-of-interest in the segmented Si-array was defined as the energy interval between $15$ and $21\,\mathrm{keV}$, corresponding to the signal of low-energy secondary electrons from radon-induced background processes, accelerated by the electric potential of the PS.

In the following, we give an overview of the background levels reached in the three measurements. We employed two generic methods to infer the level of radon-induced background (see Fig.~\ref{fig:baffle-analyse}):

\begin{itemize}
	\item a method based on identifying and counting single radon $\alpha$-decays -- the `radon spikes' discussed 
in section~\ref{sec:sub:radonbackground}. In the following, a radon spike is defined as a time interval where the 
background rate is higher than $R_{\mathrm{cut}} = 3 \cdot 10^{-2}\,$cps, which is well above the normal 
Poisson-distributed background level (see Fig.~\ref{fig:measurement-warm} and Fig.~\ref{fig:baffle-analyse}). 
We also require a specific topology of detector pixel hits: the identification of the characteristic ring-like 
pattern. The data selection criteria developed in \cite{Fraenkle:2010:PHD,RadonPaper} are used to identify 
high-energy electrons trapped in the magnetic bottle of the PS; these electrons were originated from 
shake-off processes or internal conversion accompanying $\alpha$-decay. The duration of the elevated 
background level reflects the cool-down time of the stored multi-keV electron(s) (see \cite{Fraenkle:2010:PHD} 
and Fig.~5 in \cite{nancy2} for a comparison of experimental data and MC simulation). This technique also 
allows the identification of ${}^{219}$Rn $\alpha$-decays where an internal conversion took place by 
requiring a spike duration of longer than $100\,$s (see Fig.~\ref{fig:ringcount}).
	\item a statistical method based on monitoring the mean background rate (in units of $10^{-3}\,$cps) 
between spikes in $1000\,$s time intervals. In this case the intrinsic detector background rate of 
$R_{\mathrm{det}} = (6.3 \pm 0.2)\cdot 10^{-3}\,$cps is elevated due to Rn-associated shell-reorganization 
events, where only one low-energy eV-scale electron per $\alpha$-decay is guided to the detector (see \cite{nancy} 
for a detailed discussion of this background class). We conservatively assume that cosmic-ray induced background from 
muon interactions in the vessel walls does not contribute to this background class due to the excellent 
magnetic shielding with a rather large central magnetic field $B_{\mathrm{min}}$ in the center of the PS.
\end{itemize}

Both methods yield complementary information on radon-induced background. The statistical method offers the 
advantage of better statistical precision, while the radon spike method allows the counting of single radon 
$\alpha$-decays and the inference, in principle, of the nature of the isotope undergoing $\alpha$-decay. 
The drawback of the radon spike method is the limited statistics due to the small probability of a few 
percent to produce high-energy keV-electrons via shake-off and internal conversion.   

\subsection{Background results}
\label{sec:sub:background-results}

Measurement 1 was taken with the baffle at ambient temperature shortly after its mounting. The observed 
background rate of the $120.5\,$h measurement is shown in Fig.~\ref{fig:measurement-warm}(a). For comparison with a 
previous measurement with the NEG pump but without the baffle, see Fig.~5 of \cite{RadonPaper}. Both figures show 
the generic background pattern used in our background analysis strategy: a flat background rate interrupted by rate spikes.
In this configuration, the non-elevated background periods show a Poisson-distributed background rate of 
$(10.41 \pm 0.09) \cdot 10^{-3}\,\mathrm{cps}$ (see Fig.~\ref{fig:baffle-analyse}), well above the intrinsic 
detector background rate $R_{\mathrm{det}}$. This transforms to a radon-induced singles background of 
$(4.11 \pm 0.29) \cdot 10^{-3}\,$cps (see also Tab.~\ref{tab:PS_Baffle}).

\begin{figure}[t]
  \centering
  \includegraphics[width=\textwidth]{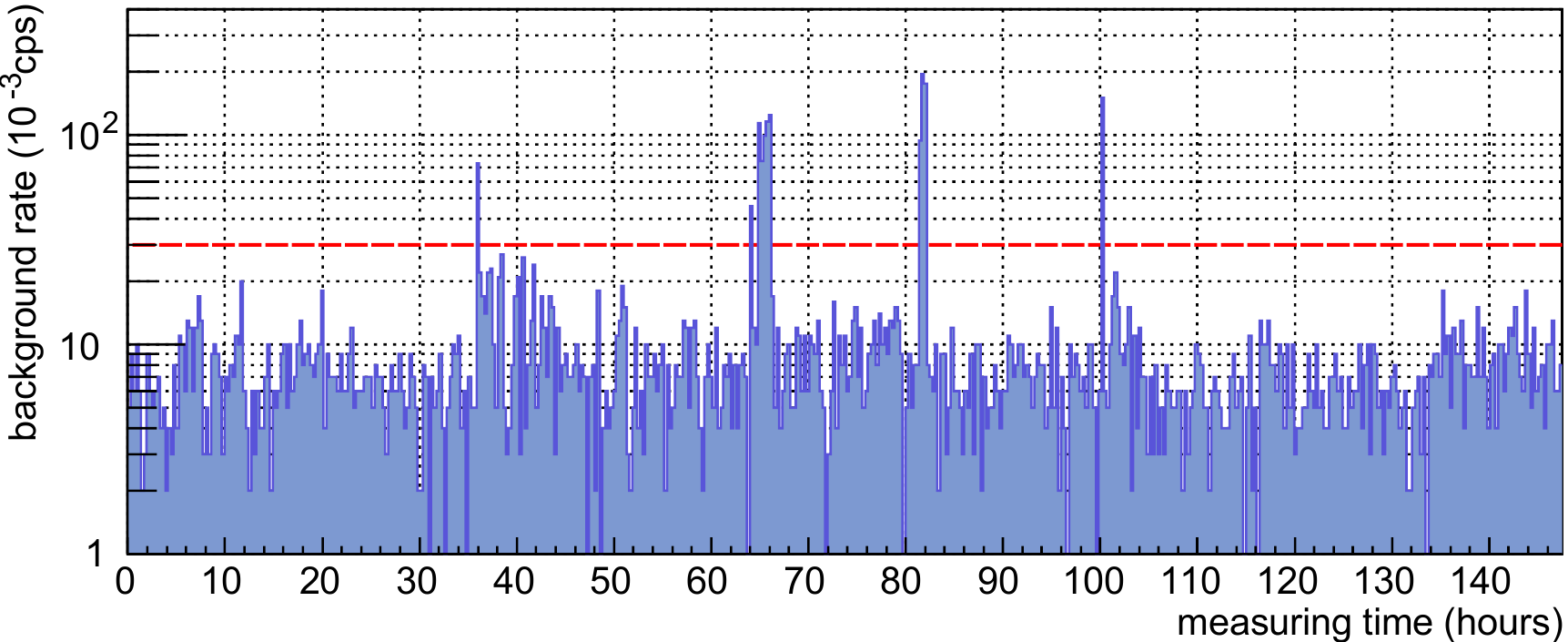}
  \caption[measurement with cold baffle]{Measurement 2: observed background rates in $1000\,$s time intervals with 
  the liquid-nitrogen-cooled baffle. The rates include the intrinsic detector background of 
  $(6.3 \pm 0.2) \cdot 10^{-3}\mathrm{cps}$. The dashed red line of $0.03\,$cps discriminates radon 
  spikes from Poisson-distributed background (see Fig.~\ref{fig:baffle-analyse}).}
    \label{fig:measurement-cold}
\end{figure}

When applying the definition of a radon spike to the data shown in Fig.~\ref{fig:measurement-warm}b, a total of 
$19$ radon spike events was observed\footnote{As shown in Fig.~\ref{fig:ringcount} and Fig.~5 of \cite{nancy2}, 
rare internal conversion processes accompanying ${}^{219}$Rn $\alpha$-decay can result in the emission of very 
high-energy electrons with a storage time exceeding the fixed time intervals of $1000\,$s shown in 
Fig.~\ref{fig:measurement-warm}. We take this into account by combining sequential time intervals with elevated 
rates above $0.03\,$cps into one single radon spike.}. This number corresponds to a rate of ($3.8 \pm 0.9$) radon 
spikes per day, which agrees within statistical uncertainties with the average background rate of ($5.4 \pm 2.4$) 
spike events per day \cite{Fraenkle:2010:PHD}, observed before the baffle was attached to the getter pump. When 
using the duration of a radon spike to discriminate high-energy electrons from internal conversion after 
${}^{219}$Rn $\alpha$-decay, $5$ out of $19$ radon spikes can be attributed to this isotope 
(see Fig.~\ref{fig:ringcount}). The remaining $14$ radon spikes with short duration stem from both 
${}^{219}$Rn and ${}^{220}$Rn.

Figure \ref{fig:measurement-cold} shows the results of measurement $2$. After turning on the liquid nitrogen supply 
to keep the baffle at LN$_2$ temperature, a reduced mean background rate of $(7.24 \pm 0.05) \cdot 10^{-3}\,\mathrm{cps}$ 
was measured in between radon spikes, close to but significantly higher than $R_{\mathrm{det}}$. The remaining 
`singles' background rate of $(0.94 \pm 0.25) \cdot 10^{-3}\,$cps is again attributed to shell reorganization 
electrons from the remaining radon $\alpha$-decays in the flux tube. This underlines the fact that the baffle 
does remove a large fraction, but not all, emanated radon atoms, as discussed in section~\ref{sec:sub:efficiency}. 
From this, we deduce a background reduction factor of $(4.4 \pm 1.2)$ for radon-induced `singles' background.

\begin{figure}[t]
  \centering
  \includegraphics[width=12cm]{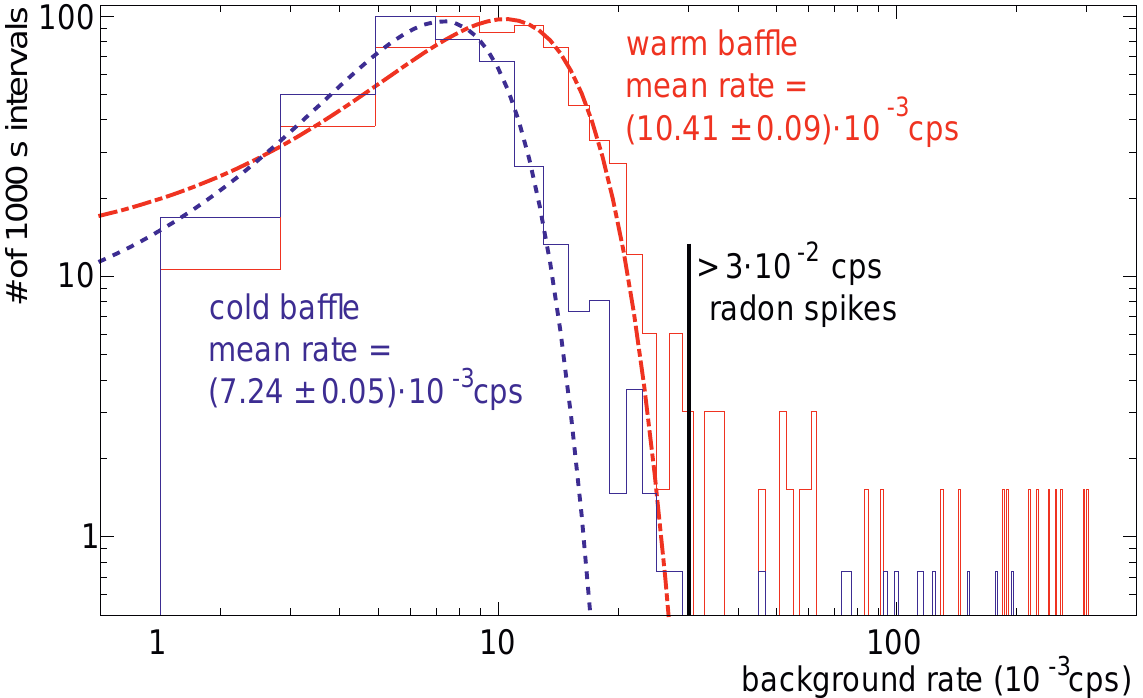}
  \caption[detailed analyses]{Distribution of observed background rates in 1000\,s long time intervals for 
  the baffle at LN$_2$ (blue histogram, fit: dashed line) and room temperature (red histogram, fit: dot-dashed line). 
  The Gaussian-fits are shown as well. Intervals with an elevated background rate of $> 0.03\,$cps are defined as 
  radon spike events, where multi-keV electrons from radon $\alpha$-decays are stored due to the magnetic bottle 
  effect. Consecutive time intervals with elevated rates are counted as a single radon spike.}
    \label{fig:baffle-analyse}
\end{figure}

Complementary information can again be obtained from the identification of individual radon spikes. 
During the measurement time of $147\,$h with the cold baffle, a total of $4$ characteristic radon spikes 
were observed (see Fig.~\ref{fig:measurement-cold}). This corresponds to a radon spike rate of 
$(0.7 \pm 0.35)$~per~day, which is substantially lower than the rate observed with a warm baffle. 
The corresponding background reduction factor of $(5.4 \pm 2.7)$ agrees well (within statistical errors) 
with the background reduction factor of $(4.4\pm1.2)$ obtained by the statistical analysis. This consistency 
gives further proof to the very good efficiency of the baffle in suppressing radon-induced background. 
Interestingly, one out of the four observed radon spikes is identified as having the origin of an 
internal conversion following ${}^{219}$Rn $\alpha$-decay. In view of the expected large suppression 
factor of ${}^{219}$Rn emanated from the NEG pump we attribute this spike to emanation of the same isotope 
from the bulk.

\begin{table}[t]
\footnotesize
\centering
\caption[measurement results]{Summary of measurement results for two different baffle operation temperatures 
(LN$_2$, room temperature) and for a `blank' run without baffle and getter. For each configuration, the 
measurement time, the overall and radon-induced singles background rate, as well as the observed number and 
rate of radon spikes are given. The radon-induced singles background rate is determined by subtracting 
the intrinsic detector background of $(6.3 \pm 0.2) \cdot 10^{-3}\mathrm{cps}$.}
\label{tab:PS_Baffle}
\begin{tabular}{l l l l l l}
\hline
measurement configuration & measurement & total singles & radon singles & number of & radon spike\\
& time (h) & rate ($10^{-3}\mathrm{cps}$) & rate ($10^{-3}\mathrm{cps}$) & radon spikes & rate (1/d)\\
\hline
1. baffle at ambient temperature & $120.5$ & $10.41 \pm 0.09$ & $4.11 \pm 0.29$ & $19$ & $3.8 \pm 0.9$\\
2. baffle cooled with LN$_2$ & $147$ & $7.24 \pm 0.05$ & $0.94 \pm 0.25$ & $4$ & $0.7 \pm 0.3$\\
3. without NEG pump & & & & & \\
and without baffle & $100$ & $7.45 \pm 0.05$ & $1.15 \pm 0.25$ & $3$ & $0.7 \pm 0.4$\\
\hline
\end{tabular}
\end{table}

\begin{figure}[t]
  \centering
  \includegraphics[width=\textwidth]{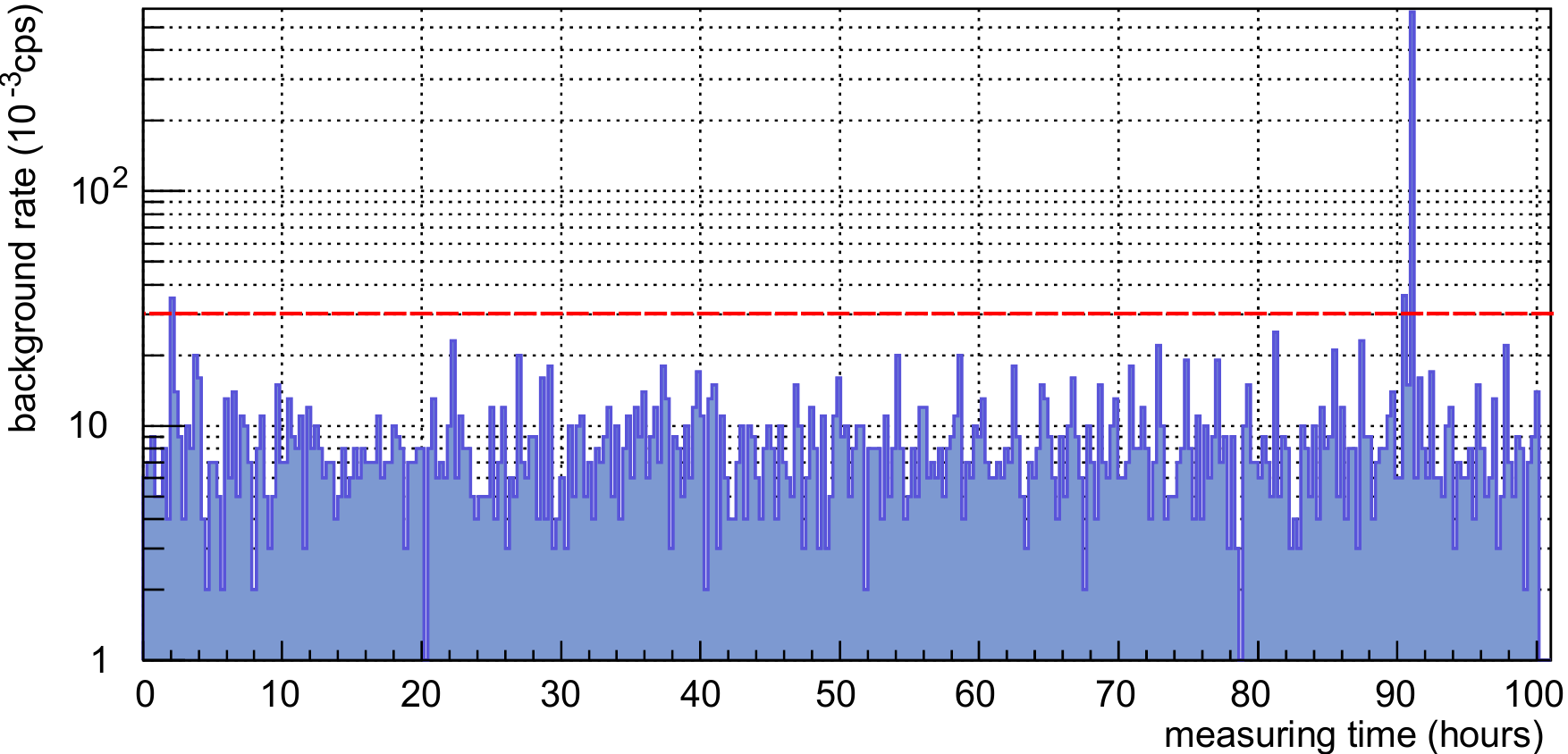}
  \caption[measurement without getter and baffle]{Measurement 3: observed background rates in $1000\,$s time
   intervals without NEG pump and without baffle. The rates include the intrinsic detector background 
   of $(6.3 \pm 0.2) \cdot 10^{-3}\mathrm{cps}$. The dashed line of $0.03\,$cps discriminates radon spikes 
   from Poisson-distributed background. The mean rate in non-elevated time periods is 
   $(7.45 \pm 0.05)\cdot 10^{-3}\,$cps. The radon spike at $t = 90.5\,$h with $634$ hits gives the 
   first experimental proof for emanation of ${}^{219}$Rn from the bulk, as only high-energy electrons 
   from internal conversion can generate more than $100$ hits per spike (see Fig.~\ref{fig:ringcount})} 
    \label{fig:blankrun}
\end{figure}

\begin{figure}[t]
  \centering
  \includegraphics[width=9.5cm]{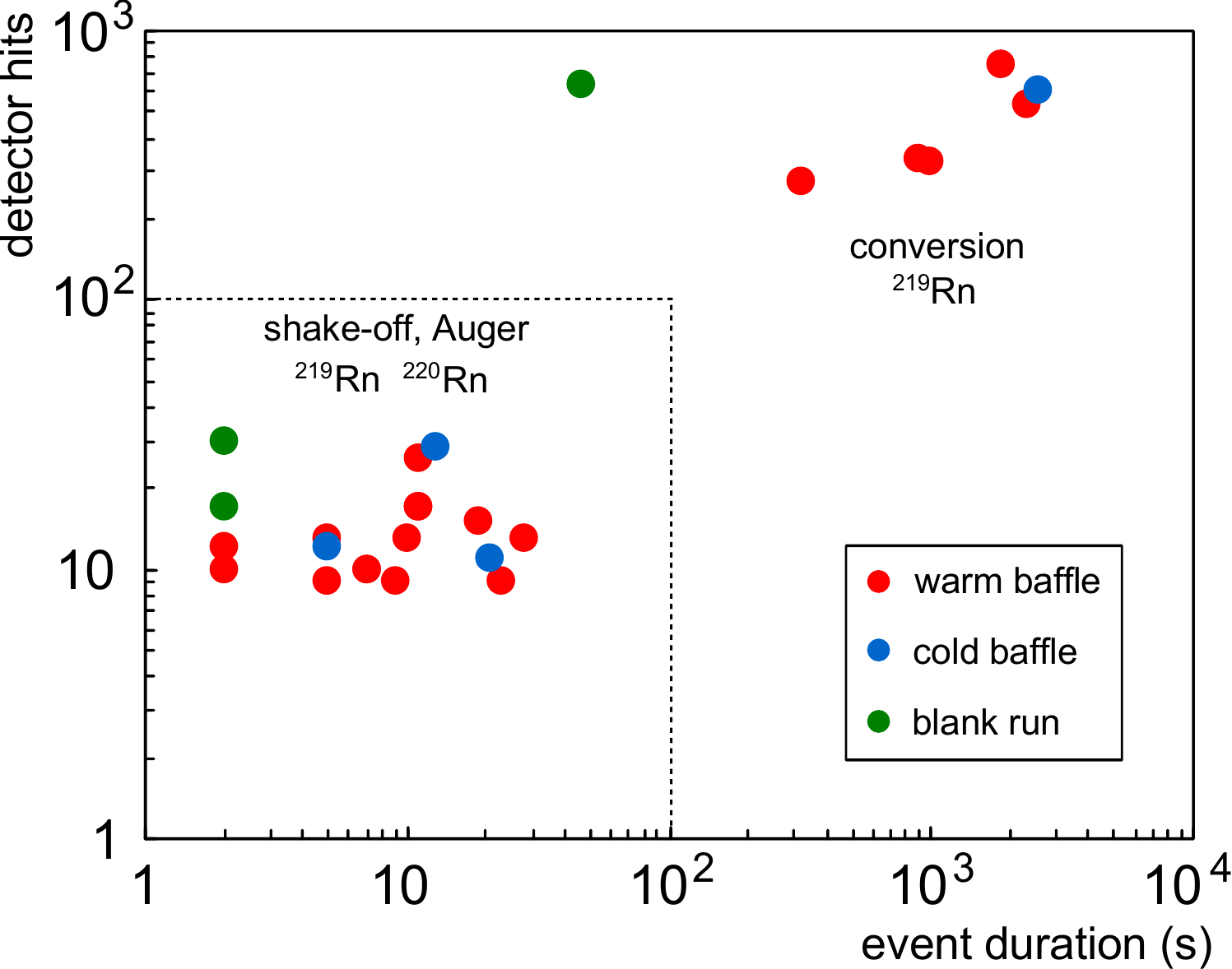}
  \caption[ring counter]{The number of detector hits as a function of the event duration for the three measurement campaigns (warm baffle = red, cold baffle = blue, blank run = green) can be used to separate internal conversion electrons from ${}^{219}$Rn $\alpha$-decay from radon spikes initiated by electrons from shake-off and Auger processes following $\alpha$-decays of both ${}^{219}$Rn and ${}^{220}$Rn. For each background class, consistent suppression factors provided by a cold baffle are obtained.} 
    \label{fig:ringcount}
\end{figure}

In the final measurement $3$, the `blank run', both the baffle and the NEG pump were removed 
\cite{RadonPaper, Fraenkle:2010:PHD} to perform a consistency check and to obtain information 
on the background contribution stemming from radon emanation from the vessel surface. 
Figure~\ref{fig:blankrun} shows the distribution of the background rate in the $100\,$h measurement. 
In the non-elevated periods a mean background rate of $(7.45\pm0.05)\cdot10^{-3}\,$cps was observed, thus 
revealing a radon-induced singles background of $(1.15\pm0.25)\cdot10^{-3}\,$cps. This level is 
slightly larger than the observed radon-induced singles rate of $(0.94\pm0.25)$ with the cold baffle, 
indicating a large suppression of ${}^{219}$Rn from the NEG pumps and a partial suppression of 
${}^{219,220}$Rn from the walls of the PS. In the complementary analysis of elevated background 
intervals, $3$ radon spikes were observed, corresponding to ($0.7 \pm 0.4$) radon ring events per day 
(see Fig.~\ref{fig:blankrun}). By removing the NEG pump and the LN$_2$ baffle, the remaining radon 
spike events were predominantly emanation of ${}^{219,220}$Rn from the inner vessel surface and 
the weld seams (see \cite{nancy2} and references therein). We note that the radon spike rate for a 
cold baffle is again smaller than in the blank run, as in the case of the statistical analysis. 
The Pre-Spectrometer measurements thus have shown that the dominant background source from 
${}^{219}$Rn emanation from the NEG pump can be efficiently blocked by the baffle system. At the same time,
the results indicate that the baffle is less efficient in removing ${}^{219,220}$Rn atoms emanated 
from the bulk. The results of all baffle measurements are summarized in Tab.~\ref{tab:PS_Baffle}. 

\section{Summary and outlook}
\label{sec:outlook}

The short-lived radon isotopes ${}^{219}$Rn and ${}^{220}$Rn which emanate from the porous NEG 
strips and the surface of the spectrometer are a serious source of background in electrostatic 
retarding spectrometers due to the emission of electrons with energies up to 200 keV. As a result 
of the inherent magnetic trap formed by the MAC-E-filter, single $\alpha$-decay processes of these 
isotopes result in enhanced levels of background over an extended period of time, adding a 
non-Poissonian background component \cite{NuclearDecayPaper} to the cosmic-ray induced background. 
Thus, radon emanation seriously limits the neutrino mass sensitivity of KATRIN if no countermeasures 
are taken. Due to the fact that the emanating radon atoms are neutral, the highly efficient magnetic 
and electrostatic shielding against charged particles from the vessel walls is ineffective. This deficiency
calls for additional active and passive background suppression techniques.

Here we have reported on the implementation of a passive method, in form of a LN$_2$-cooled baffle 
that traps radon emanated from the NEG pump and the inner vessel walls. By comparing measurements 
with a LN$_2$-cooled baffle to measurements with a warm baffle, an overall singles background reduction 
factor of $(5.4 \pm 2.7)$ has been observed. By identifying individual radon spikes from high-energy 
trapped electrons, a comparable background reduction factor of $(4.4 \pm 1.2)$ was observed, giving a 
first and consistent proof-of-principle of this novel background reduction technique in MAC-E filters.

The Main Spectrometer ($10^{-11}\,\mathrm{mbar}$) was originally equipped with a scaled-up version 
of the NEG pump described above, with $3\,$km of St707 getter strips installed. By extrapolating the results of 
the PS tests reported here and in \cite{nancy2}, the expected radon-induced background rate due to 
the NEG material and the large surface of the Main Spectrometer was estimated to be of the order of 
$1\,\mathrm{cps}$ \cite{Fraenkle:2010:PHD, NuclearDecayPaper, Mertens:2012:PHD}. This value was 
confirmed by the commissioning measurements (\cite{Goerhardt:2014:PHD}, \cite{LRT2017}). As such 
a large background rate would significantly reduce the sensitivity of KATRIN on 
$m_{\nu}$ \cite{NuclearDecayPaper, Mertens:2012:PHD, Wandkowsky:2013:PHD}, this background class 
has to be further reduced by more than two orders of magnitude.

Based on the encouraging results of this paper, a large-scale baffle system to reduce 
radon-induced background in the Main Spectrometer has been designed and optimized along the design 
criteria outlined above. In order to reach a similar pumping speed for radon in the 
much larger Main Spectrometer (690\,m$^2$), the cold surface would have to be 28 times larger 
(required: 5.5\,m$^2$ cold surface area, actual design: 6.6\,m$^2$). In addition, one has to take 
into account the 7 times longer time of flight between two hits of the wall, which would increase 
the decays in flight by approximately the same factor. For radon-induced background rates in the PS of up to 
$2\cdot 10^{-3}$\,cps, the scaled-up rate in the Main Spectrometer is still close 
enough to the limit proposed in the KATRIN Design Report \cite{KATRINDesignReport}. 
For the Main Spectrometer a modified design was adopted for the baffles. It retains a 
sufficiently large effective pumping speed for hydrogen and tritium (40\% of the value without baffles) 
by abandoning the central disk in the PS baffle design \cite{lit:SDSvacuum16}. 
The large Main Spectrometer baffles were made OFHC copper \cite{RadonAdsorption}. 
In principle, the baffle operation and radon adsorption times $\tau_{\mathrm{des}}$ could be 
further improved by gold-plating the V-shaped copper panels or by annealing the base material 
(copper) at a temperature of $\sim 1000\,$K \cite{RadonAdsorption}. A detailed description of 
this large-scale baffle and its performance is given in \cite{Goerhardt:2014:PHD, MSradonsimu2017}.

In addition to the passive background reduction via the technique of a liquid-nitrogen-cooled 
copper baffle discussed here, further active suppression methods of background induced by 
magnetically stored electrons were investigated, such as a cyclic application of 
an electric dipole \cite{Hilk:2017:Thesis} in combination with a magnetic pulse 
\cite{Wandkowsky:2013:PHD, Behrens:2017:Thesis, arXivBehrens2018}, and electron cyclotron 
resonance \cite{ECRpaper, Mertens:2012:PHD}. 
The method of removing emanated Rn atoms from the active flux tube of a MAC-E filter is studied 
in more detail in measurements with the Main Spectrometer \cite{Goerhardt:2014:PHD, LRT2017}.
By combining all the background reduction techniques, a staged approach to reach conditions 
where the spectrometer is essentially free of radon-induced background is at hand 
for the KATRIN neutrino mass measurements.


\section*{Acknowledgements}

This work has been supported by the Bundesministerium f\"ur Bildung und Forschung (BMBF) with project numbers 05A11VK3 and 05A11PM2. The authors acknowledge the excellent support of the Pre-Spectrometer operation by the KATRIN technical staff, in particular by H.~Skacel, B.~Bender and A.~Beglarian. S.M. acknowledges support by a Feodor Lynen Research fellowship of the Alexander von Humboldt-Foundation, as well as support by the Helmholtz Association. S.G., F.M.F. and S.M. would like to thank KHYS at KIT for support. Special thanks go to Robert Eichler for very helpful discussions and to Xueli Luo for his accurate calculations.


\bibliographystyle{aipnum4-1}
\bibliography{references}

\end{document}